\UseRawInputEncoding
\documentclass[pre,preprint,preprintnumbers,amsmath,amssymb]{revtex4}
\usepackage{graphics,epsfig,epstopdf,stmaryrd}
\usepackage{textcomp}
\usepackage{bm}

\newcommand{\mE}{\mathcal{E}}
\newcommand{\mA}{\mathcal{A}}
\newcommand{\mI}{\mathcal{I}}

\newcommand{\lcos}{\langle\cos(j\varphi)\rangle}
\newcommand{\rf}{Ref.~\cite}
\begin {document}
\title{Nonlinear adiabatic electron plasma waves. I. General theory and nonlinear frequency shift.}
\author{Mikael Tacu }
\author{Didier B\'enisti}
\affiliation{ CEA, DAM, DIF F-91297 Arpajon, France and Universit\'e Paris-Saclay, CEA, LMCE, 91680 Bruyres-le-Ch\^atel, France.}
\begin{abstract}
This paper provides a complete self-consistent nonlinear theory for electron plasma waves, within the framework of the adiabatic approximation. The theory applies whatever the variations of the wave amplitude, provided that they are slow enough, and it is also valid when the plasma is inhomogeneous and non stationary. Moreover, it accounts for: (i) the geometrical jump in action resulting from separatrix crossing; (ii) the continuous change in phase velocity making the wave frame non-inertial; (iii) the harmonic content of the scalar potential ; (iv) a non-zero vector potential ;  (v) the transition probabilities from one region of phase space to the other when an orbit crosses the separatrix ; (vi) the possible change in direction of the wavenumber. The relative importance of each of the aforementioned effects is discussed in detail, based on the derivation of the nonlinear frequency shift. This allows to specify how the general formalism may be simplified, depending on the value of the wavenumber normalized to the Debye length. Specific applications of our theory are reported on the companion paper. 
\end{abstract}
\maketitle
\section{Introduction}
\label{0}
In this article, and in the companion paper~\cite{benisti20II}, the term electron plasma wave (EPW) refers to any wave that essentially results from the electron motion, in a plasma with no magnetic field but that induced by the wave itself. The EPW is assumed to be electrostatic in the linear regime but, as shown in Paragraph~\ref{IID}, in the nonlinear regime and in two or three dimensions, a magnetic field necessarily builds up. Hence, our definition for EPW's is quite general, and it clearly does not restrict to Langmuir waves~\cite{langmuir}. 

One major result of this paper is the derivation of the nonlinear electron distribution function, when the nonlinearity is essentially due to electron trapping in the wave trough. This requires a self-consistent description of the EPW scalar and vector potentials, and of the EPW nonlinear frequency, which are the two other major results of the article. 

Electron trapping often leads to the first nonlinear regime of plasma instabilities, either self-consistent such as beam-plasma instabilities~\cite{bret1,grem1}, or externally driven such as stimulated Raman scattering (SRS)~\cite{kruer}. Indeed, as is well-known, trapping leads to the saturation of the beam-plasma instability, when the beam is cold enough~\cite{bret1,oneil71}. It also strongly reduces the Landau damping rate of a driven plasma wave~\cite{oneil65,benisti09,yampo}, leading to the so-called regime of kinetic inflation of SRS~\cite{vu,montgomery}. Consequently, as discussed in Section~\ref{I}, the results of this paper are directly relevant to address the nonlinear stage of plasma instabilities. 

In the trapping regime, the wave electric field reads,
\begin{equation}
\label{1}
E(\bm{x},t)Ê=\mE\left[\bm{x},t,\varphi(\bm{x},t)Ê\right],
\end{equation}
with $\mE(\varphi+2\pi)=\mE(\varphi)$. From the eikonal, $\varphi$, one introduces the wavenumber, $\bm{k}=\bm{\nabla}\varphi$, and the wave frequency, $\omega=-\partial_t\varphi$. Then, in the trapping regime, the space and time variations of $\varphi$ are much faster than those of $\bm{k}$, $\omega$, and $\mE$ (at fixed values of $\varphi$). 

\begin{figure}[!h]
\centerline{\includegraphics[width=8.6cm]{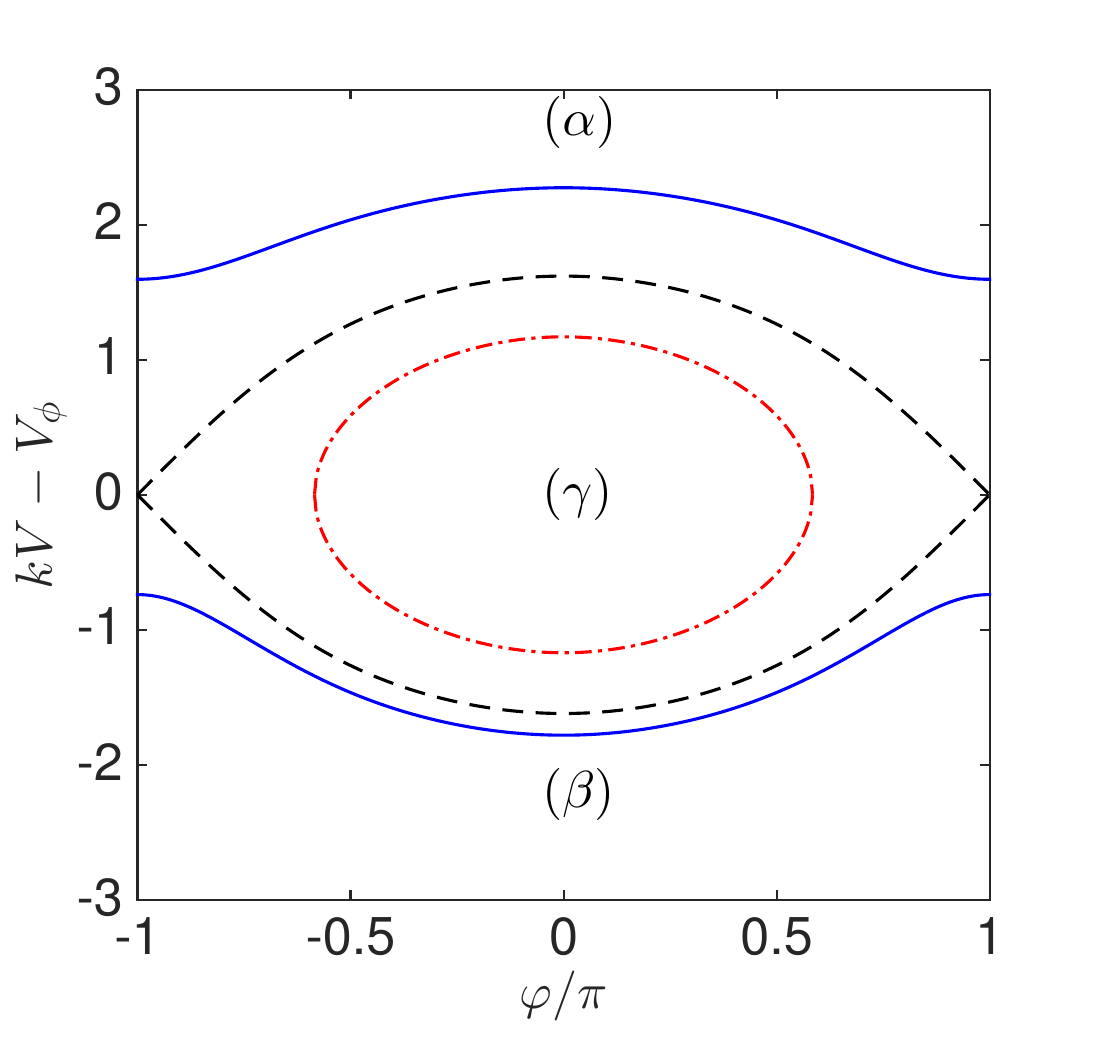}}
\caption{\label{f1} (Color online) The blue solid lines show two different untrapped orbits, the red dashed-dotted line shows a trapped orbit, while the black dashed line shows the separatrix, when $\Phi$ defined by Eq.~(\ref{Phi}) is $\Phi=\Phi_1\cos(\varphi)+\Phi_2\cos(2\varphi)+\Phi_3\cos(3\varphi)$, with $\Phi_1=0.645$, $\Phi_2=-7.16\times10^{-2}$ and $\Phi_3=1.04\times10^{-2}$.}
\end{figure}

Replacing $\mE(\bm{x},t,\varphi)$ with $\mE(\bm{x}_0,t_0,\varphi)$, where $\bm{x}_0$ and $t_0$ are constants, yields the so-called frozen dynamics. The corresponding frozen orbits are of two different kinds. As shown in  Fig.~\ref{f1}, the untrapped frozen orbits in regions $(\alpha)$ and $(\beta)$ of phase space are unbounded in $\varphi$, unlike the trapped orbits in region $(\gamma)$. The aforementioned regions of phase space are delimited by a curve called the separatrix, and represented by the black dashed line in Fig.~\ref{f1}. Then, nonlinear wave-particle interaction is in the trapping regime if the maximum width, in velocity, of the separatrix is much larger than the width in phase velocity of the Fourier spectrum of $E$ (while, in the opposite limit, one would rather expect quasilinear diffusion~\cite{benisti97}). 

Clearly, the frozen orbits are $2\pi$-periodic in $\varphi$ so that, for the frozen dynamics, the phase-space is topologically equivalent to a cylinder. On the cylinder, all frozen orbits, either trapped or untrapped, are closed, and are considered as such. Using this convention lets us define the dynamical action, $I$, as the phase-space area of a frozen orbit divided by $2\pi$. When the space and time variations of the dynamics are slow enough, the action is known to be a well preserved quantity. Indeed, as shown in Refs.~\cite{neishtadt,benisti16}, the action of orbits which remain in region $(\gamma)$ is nearly conserved, $d_tI^{(\gamma)}\approx 0$. As for untrapped orbits, $d_tI^{(\alpha,\beta)}\approx-k^{-1}\eta\partial_{x_r} H_0\vert_I$, where $x_r$ is along the direction of the wavenumber, $H_0$ is the Hamiltonian for the electron dynamics, and $\eta=1$ in region $(\alpha)$ while $\eta=-1$ in region $(\beta)$. When the latter equations for the time evolution of the actions are fulfilled, we call the corresponding EPW an ``adiabatic electron plasma wave''. In the remainder of this article, and in the companion paper~\cite{benisti20II}, we restrict to such waves. Now, although the results on the time variations of the action are usually well-known, when an EPW may be considered as adiabatic and how the results on the action may actually be applied to EPW's is usually much less known. Therefore, we find it useful to quickly recall it in Section~\ref{I}.

When the dynamics does not explicitly depend on space, i.e., when the wave electric field reads $E=\mE(t,\varphi)$, the action is an adiabatic invariant, \textit{but only within each region of phase space}. $I^{(\alpha,\beta,\gamma)}\approx Const.$, which does not mean that the action distribution function may be derived by assuming $I\approx Const.$ Indeed, let $V_\phi=\omega/k-eA_0/m$, where $A_0$ is the wave vector potential. Then, as shown in Paragraph~\ref{IIB}, the action changes by a multiple of $V_\phi^*/k^*$ any time an orbit crosses the separatrix, where $V_\phi^*$ and $k^*$ are the values assumed by $V_\phi$ and $k$ when separatrix crossing occurred.  Consequently, the action distribution function depends on the whole history of $V_\phi/k$. This has been underlined in Refs.~\cite{benisti17I,benisti17II}, where it has been shown that the wave frequency usually was a non-local function of its amplitude. 

Separatrix crossing lets an orbit move from one region of phase space into the two other ones, with specific transition probabilities. These need to be correctly estimated in order to derive the action distribution function. Furthermore, the very expression of the action depends on the space profile of the scalar potential, while the value of $V_\phi$ depends on the wave frequency and vector potential. All these quantities need to be derived self-consistently, together with the action distribution function, in order to provide a complete description of nonlinear adiabatic EPW's. An additional complexity arises in the inhomogeneous situation when the wave frequency is space-dependent. Indeed, the consistency relation, $\partial_t\bm{k}=-\bm{\nabla}\omega$, usually entails a change in the $\bm{k}$-direction, which also needs to be derived self-consistently. In this paper, we show how to address all the aforementioned issues. Moreover, we provide an explicit derivation of the nonlinear frequency shift, $\delta \omega$ (defined as the difference between the nonlinear frequency, $\omega$, and its linear limit $\omega_{\rm{lin}}$), when the wave is homogeneous and when it keeps growing. Indeed, only in this situation is $\delta \omega$ a local function of the wave amplitude. We also explain how the general theory may be simplified, e.g., by  assuming a sinusoidal scalar potential or by neglecting the change in the $\bm{k}$-direction, depending on the specific physics situation which is addressed. 

Several authors previously derived the adiabatic distribution function and nonlinear wave frequency, e.g., in Refs.~\cite{benisti17II,benisti07,benisti08,dewar,krapchev,russell,lindberg,dodin}. However, we are not aware of any previous publication addressing the effect entailed by the change in the $\bm{k}$-direction. Moreover, all the aforementioned articles, except Ref.~\cite{benisti17II}, were only valid for a homogeneous growing wave. Furthermore, only in Ref.~\cite{lindberg} were the harmonic content of the scalar potential and the vector potential of the EPW self-consistently derived, only in Refs.~\cite{benisti17II,benisti07,benisti08} was the non-locality of the distribution function accounted for, and only in Ref.~\cite{benisti17II} were the transition probabilities correctly estimated. Hence, to the best of our knowledge, no previous theory addressed the very general situation considered in this paper. However, unlike in Ref.~\cite{benisti08}, we do not account for the effect of an external drive when deriving the nonlinear frequency shift, $\delta \omega$. Indeed, we did not find necessary to reproduce the derivation of Ref.~\cite{benisti08} since no new result is to be expected compared to what has already been published in the latter paper.

This article is organized as follows. Section~\ref{I}~quickly recalls when the adiabatic theory may be used to derive the nonlinear properties of electron plasma waves. In Section~\ref{II}, we detail the derivation of the electron distribution function, and of the EPW nonlinear frequency shift. In Section~\ref{III}, we provide explicit values for the nonlinear frequency shift, $\delta \omega$, when the EPW is uniform and keeps growing in an initially Maxwellian plasma. Moreover, from the values found for $\delta \omega$, we discuss how the general theory may be simplified. 
Section~\ref{IV} summarizes and concludes the article.

\section{Applicability of the adiabatic theory to nonlinear electron plasma waves}
\label{I}

When the electric field does not explicitly depend on space, i.e., when it reads $E(\bm{x},t)=\mE[t,\varphi(\bm{x},t)]$, and when the time periods of frozen orbits are much smaller than any other timescale of the problem, it is well-known that the action $I$ is an adiabatic invariant~\cite{neishtadt,arnold}. Moreover, it is also well-known that the time period of {\em{all}} orbits need not be small for the action to remain nearly constant. Indeed, although the period on the separatrix is infinite, the adiabatic approximation is still valid to derive the change in action entailed by separatrix crossing, which mainly consists in a geometrical jump~\cite{cary}.  However, these mathematical results do not really tell us how to apply the adiabatic approximation, $I^{(\alpha,\beta,\gamma)}\approx Const.$, to EPW's. Let us quickly recall here the conditions for the applicability of the latter approximation, already reported in Ref.~\cite{benisti07}. 

One may make use of the adiabatic approximation to derive the EPW dispersion relation whenever,
\begin{equation}
\label{adia}
\frac{\Gamma}{kv_{\rm{th}}} \alt 0.1.
\end{equation}
In Eq.~(\ref{adia}), $k$ is the EPW wavenumber, $v_{\rm{th}}$ is the typical width in velocity of the unperturbed distribution function (for a Maxwellian, this would just be the thermal speed), and $\Gamma$ is the typical rate of variation of the field amplitude, as seen by an electron,
\begin{equation}
\label{2}
\Gamma \equiv \left.\frac{d\ln\left\vert\mE\left[x(t),t,\varphi\right]\right\vert}{dt}\right\vert_{\varphi=Const.}.
\end{equation}
 Note that the periods of frozen orbits do not explicitly enter Eq.~(\ref{adia}). They only have to be accounted for when it comes to deriving the variations of the wave amplitude. Indeed, for a growing wave fulfilling Eq.~(\ref{adia}), the adiabatic approximation becomes accurate to describe the nonlinear evolution of the wave amplitude only once,
\begin{equation}
\label{adia2}
\Gamma T_B\alt 1/2,
\end{equation}
where $T_B$ is the typical period of a frozen trapped orbit. More precisely, $T_B=2\pi/\sqrt{ekE_{\max}/m}$~\cite{note}, $-e$ being the electron charge, $m$ its mass and $E_{\max}$ the maximum value of the electric field over one space period. When the EPW does not keep growing, Eq.~(\ref{adia2}) may be replaced by $\int dt/T_B[E_{\max}(t)]\agt1$. 

When the electric field explicitly depends on space, so that the dynamics is not exactly periodic, the conditions Eqs.~(\ref{adia}) and (\ref{adia2}) still apply but, as already indicated in the Introduction, the adiabatic approximation now reads, $I^{(\gamma)}\approx Const.$, $d_tI^{(\alpha,\beta)}\approx -\eta k^{-1}\partial_{x_r}H_0\vert_I$. 

Eq.~(\ref{adia}) is very well fulfilled for SRS-driven EPW's in a laser-fusion plasma, so that the adiabatic approximation applies very well to such waves. Indeed, it has been shown in Ref.~\cite{friou} that the space-averaged adiabatic distribution function was very close to that derived numerically from a Vlasov simulation of SRS. The adiabatic nonlinear frequency shift has also been compared against results from 1-D Vlasov simulations of SRS in Ref.~\cite{benisti08}, and shown to be quite accurate. Moreover, in Refs~\cite{brama,SRS3D,ppcf}, the adiabatic theory has been successfully used to describe the nonlinear evolution of SRS, including the regime of kinetic inflation, whenever Eq.~(\ref{adia2}) was fulfilled. Nevertheless, for small amplitude waves, the use of the adiabatic approximation would be inadequate because it would not allow for Landau damping. However, the way to overcome this difficulty, to address SRS for small amplitude waves and to study the transition between the linear and the inflation regime, has been explained in Refs.~\cite{benisti15,benisti10}, and will be quickly recalled in the companion paper~\rf{benisti20II}. 

As for the cold beam-plasma instability, it cannot be addressed by making use of the adiabatic approximation because Eq.~(\ref{adia}) is not satisfied for the value $v_{\rm{th}}$ of the beam distribution function. Yet, the concept of adiabaticity is still useful, in some sense, in order to address the nonlinear growth and saturation of this instability, as already shown in Refs.~\cite{action,belfast}. Moreover, as will be discussed in a forthcoming publication, the results obtained in this article are very useful to a description of nonlinear EPW's that goes beyond the adiabatic approximation.

\section{General derivation of the adiabatic distribution function and nonlinear frequency shift}
\label{II}
\subsection{Hypotheses and notations}
\label{IIA}
Henceforth, we only consider situations when Eq.~(\ref{adia}) is fulfilled. For the sake of clarity, the theory is first derived for a one-dimensional (1-D) geometry, assuming a constant direction for the wavenumber. Such a 1-D approach often captures the essential features of the nonlinear physics. The impact of the wavenumber rotation on the electron distribution function is explicitly addressed in the Appendix~\ref{C}, together with the range of validity of the 1-D approximation. Moreover, one important 3-D effect, the build-up of a magnetic field, is discussed in Section~\ref{IID} from a straightforward generalization of the 1-D results. 

The electric field, $E$, is that of the sole electron plasma wave, which does not necessarily mean that the ions are assumed to be motionless. Indeed, as argued in Ref.~\cite{benisti16}, in many situations the ion and electron timescales decouple, so that the effect of mobile ions only enters in the slow evolution of the action distribution function of untrapped electrons.

Since we make use of the adiabatic approximation, we derive here the distribution function in action-angle variables, where the angle, $\theta$, is canonically conjugated to the action, $I$. The distribution function, $F(\theta,I,t)$, is normalized in such a way that,
\begin{equation}
\int F[\theta(x,t),I,t]d(kI)=n_e(x,t),
\end{equation}
where $n_e(x,t)$ is the electron density. As discussed in Appendix~\ref{A}, in the adiabatic regime, $F$ may be approximated by,
\begin{equation}
\label{3}
F[\theta(x,I),I,t] \approx n_e(x,t) f(I,x,t),
\end{equation}
where
\begin{equation}
\label{7}
f(I,x,t)=\frac{k}{2\pi n_e(x,t)}Ê\int_{x-\pi/k}^{x+\pi/k} F[\theta(x',I),I,t] dx'.
\end{equation}
Note that the normalization of $f$ is such that, 
\begin{equation}
\label{norma}
\int f(I,x,t)d(kI)=1.
\end{equation}
As shown in Paragraph~{\ref{IIC}, the choice Eq.~(\ref{norma}) allows to easily express $f$ in terms of the unperturbed velocity distribution function, normalized to unity.


\subsection{Hamiltonian for the wave-particle interaction}
\label{IIB}
Henceforth, we assume that the wave electric field may be decomposed as a sum of harmonics,
\begin{equation}
\label{E}
E=E_0+\sum_{j\geq1} E_j \sin(j\varphi+\delta \varphi_j),
\end{equation}
such that, when $E_1\rightarrow 0$, $E_j/E_1\rightarrow0$ if $j\neq1$. The zeroth harmonic of the electric field, $E_0$, derives from a vector potential (which is obvious when the amplitudes of the harmonics only depend on time), denoted by,
\begin{equation}
\label{A0}
A_0(x,t)=-\int_0^tE_0(x,u)du.
\end{equation}
Hence, in general, the wave cannot be considered as purely electrostatic (see the end of Paragraph~\ref{IID}). Its electric field does not derive from a scalar potential, only $E-E_0$ does. Consequently, let us introduce $\phi(\varphi,t)$ such that,
\begin{eqnarray}
\label{phi}
\phi(\varphi,t)&=&-(e/km)\int [E(\varphi,t)-E_0(\varphi,t)]d\varphi\\
\label{phi2}
&\equiv&\sum_{j\geq1}\phi_j(\varphi,t)\cos(j\varphi_j+\delta\varphi'_j).
\end{eqnarray}
Then, the Hamiltonian for the electron dynamics (only accounting for the effect of the electric field $E$) is easily shown to be (see~\rf{benisti16}),
\begin{equation}
\label{H0}
H_0=H-\frac{V_\phi^2}{2}+e^2A_0^2/2m^2,
\end{equation}
where\begin{equation}
\label{vphi}
V_\phi=\omega/k-eA_0/m,
\end{equation}
and where,
\begin{equation}
\label{H}
H=\frac{(kV-V_\phi)^2}{2}-\phi(\varphi,t).
\end{equation}
In Eq.~(\ref{H}), the dynamical variable canonically conjugated to $\varphi$ is $V=(v-eA_0/m)/k$, where $v$ is the electron velocity. 

Now, by making use of the adiabatic approximation, one may only derive the wave dispersion relation at zeroth order in the variations of the fields amplitudes. This means that one may account for their space and time variations, but not for their space and time derivatives. Then, within this approximation, and as shown in Appendix~\ref{A}, one may choose $\delta \varphi_j=0$ in  Eq.~(\ref{E}), $\delta \varphi'_j=0$ in the expression for $\phi$, and approximate $\phi_j$ by,
\begin{equation}
\label{phij}
\phi_j=eE_j/jmk.
\end{equation}
These approximations will be used throughout the remainder of this paper.

\subsection{Derivation of the action}
\label{IIB2}
$V$ and $\varphi$ being canonically conjugated, the action is,
\begin{equation}
\label{iu}
I=\frac{1}{2\pi}\oint Vd\varphi,
\end{equation}
where the integral is calculated along a frozen orbit, located about a given $\varphi_0\equiv \varphi(x_0,t_0)$. Hence, $V$ in Eq.~(\ref{iu}) is derived from Eqs.~(\ref{H0})~and~(\ref{H}) by freezing $V_\phi$, $A_0$ and the harmonics amplitudes, $\phi_j$, at the values they assume when $x=x_0$ and $t=t_0$. Denoting by $V_{\phi_f}$, $\phi_f$ and $A_{0_f}$ the corresponding frozen values yields,
\begin{equation}
\label{igeneral}
kI=\frac{1}{2\pi}\oint \left[V_{\phi_f}\pm \sqrt{2\left(H_0+\frac{V_{\phi_f}^2}{2}+\phi_f-\frac{e^2A_{0_f}^2}{2m^2}\right)}\right]d\varphi.
\end{equation}
In the integral of the right-hand side of Eq.~(\ref{igeneral}), $H_0$, $V_{\phi_f}$ and $A_{0_f}$ are constant, so that $H=H_0+V_\phi^2/2-e^2A_0^2/2m^2$ is also a constant. Then, $I$ assumes exactly the same expression as when the electron dynamics derives from $H$ instead of $H_0$. 

Using the symmetry of frozen orbits with respect to the $V$-axis, Eq.~(\ref{igeneral}) reads, respectively for orbits lying in regions $(\alpha)$ and $(\beta)$ of phase space (see Fig.~\ref{f1}),
\begin{eqnarray}
\label{iua}
kI_u^{(\alpha)}&=&\frac{1}{\pi} \int_{0}^{\pi} \sqrt{2\left[H+\phi(u)\right]} du+V_\phi, \\
\label{iub}
kI_u^{(\beta)}&=&\frac{1}{\pi} \int_{0}^{\pi}  \sqrt{2\left[H+\phi(u)\right]} du-V_\phi,
\end{eqnarray}
where we henceforth use the notations, $\phi(u)\equiv\phi_f(u+\varphi_0,t_0)$ and $V_\phi\equiv V_\phi(\varphi_0,t_0)$.

As for trapped orbits, in order to avoid a jump by a factor of two compared to $I_u$, we do not use the general definition, Eq.~(\ref{igeneral}), but we rather define,
\begin{equation}
\label{it}
I_t^{(\gamma)}=\frac{1}{4\pi} \oint Vd\varphi=\frac{1}{\pi k} \int_{0}^{\varphi_{\max}} \sqrt{2\left[H+\phi(u)\right]}du,
\end{equation}
where $\varphi_{\max}>0$ is such that~$H+\phi(\varphi_{\max})=0$.  

From Eqs.~(\ref{iua})-(\ref{it}), for the same value of the action, an orbit may lie in region $(\alpha)$, $(\beta)$ or $(\gamma)$. Consequently, one needs to decompose the normalized distribution function, $f(I)$ defined by Eq.~(\ref{7}), into 3 different distribution functions, $f_\alpha(I)$,  $f_\beta(I)$,  and $f_\gamma(I)$,  corresponding to each of the respective region of phase space.

\subsection{Distribution function and nonlinear frequency shift for a uniform plasma and wave amplitude}
For the sake of clarity, we first derive the electron distribution function and wave nonlinear frequency shift when the plasma and wave amplitude are uniform. 
\label{IIC}
\subsubsection{Nonlinear distribution function}
\label{IIC1}
From Eqs.~(\ref{iua})~and~(\ref{iub}), it is clear that, when the wave amplitude is zero, $k_0I_u^{(\alpha)}=v_0$ and $k_0I_u^{(\beta)}=-v_0$, where $v_0$ is the unperturbed electron velocity and $k_0$ is the initial wavenumber. Moreover, when the plasma and the wave amplitude are uniform, the action is conserved, provided that no separatrix crossing has occurred. Then, using the convention that the distribution function has to be positive, we find that \textit{for orbits that have never crossed the separatrix},
\begin{eqnarray}
f_\alpha(I)=f_0(k_0I), \\
f_\beta(I)=f_0(-k_0I),
\end{eqnarray}
where $f_0$ is the unperturbed velocity distribution function, normalized to unity. However, it is clear from Eqs.~(\ref{iua})-(\ref{it}) that separatrix crossing entails a change in the action. For example, if an orbit moves from region $(\alpha)$ to region $(\gamma)$, its action changes by $-V_\phi^*/k^*$, where $V_\phi^*$ and $k^*$ are, respectively, the values assumed by $V_\phi$ and $k$ \textit{when separatrix crossing has occured}. 

The detailed derivation of the change in the action distribution function entailed by separatrix crossing may be found in Ref.~\cite{benisti17I}. The results are for a sinusoidal wave, but they may be straightforwardly generalized by replacing $v_{tr}$ in Eqs.~(92)-(97) of Ref.~\cite{benisti17I} with $\mA_s$, where $4\pi\mA_s$ is the area enclosed by the separatrix (in velocity). Let us recall here the main results:
\begin{enumerate}
\item An orbit crosses the separatrix from region $(\alpha)$ whenever $kI_u^{(\alpha)}=\mA_s+V_\phi$ while $(\mA_s+V_\phi)$ is increasing, from region $(\beta)$ whenever $kI_u^{(\beta)}=\mA_s-V_\phi$ while $(\mA_s-V_\phi)$ is increasing, and from region $(\gamma)$ whenever $kI_t^{(\gamma)}=\mA_s$ while $\mA_s$ is decreasing. 
\item Each time an orbit leaves one region of phase space, the probability that it ends up in either one of the two other regions depends on $\vert dV_\phi/d\mA_s\vert$. 
\item The most simple, and most frequent situation, is when $\vert dV_\phi/d\mA_s\vert<1$. In this case: 
\item[-]When $\mA_s$ decreases, orbits from region $(\gamma)$ move into regions $(\alpha)$ and $(\beta)$, and,
\begin{eqnarray}
\label{df2}
f^>_\alpha(I)&=&f^<_\gamma(I-V_\phi^*/k^*)/2, \\
\label{df3}
f^>_\beta(I)&=&f^<_\gamma(I+V_\phi^*/k^*)/2, 
\end{eqnarray}
where the superscript, $^<$, means ``before separatrix crossing'', the superscript $^>$, means ``after separatrix crossing''.
\item[-]When $\mA_s$ increases, orbits from regions $(\alpha)$ and $(\beta)$ move into region $(\gamma)$, and
\begin{equation}
\label{df}
f^>_\gamma(I)=f^<_\alpha(\mI_+)\frac{d\mI_+}{dI}+f^<_\beta(\mI_-)\frac{d\mI_-}{dI},
\end{equation}
where $\mI_\pm=I\pm V_\phi^*/k^*$.
\end{enumerate}
Eqs.~(\ref{df2})-(\ref{df}) yield the fraction of electrons whose action lies between $I$ and $I+\delta I$. Although these equations are quite accurate, as shown in Ref~\cite{benisti17I}, there are caveats in using them to derive averaged quantities, especially as regards the contribution from trapped electrons. Indeed, as discussed in  Ref.~\cite{DNC}, in spite of action conservation, the actual distribution function for trapped electrons, which we denote by $\mathcal{F}_\gamma$, depends on $\theta$ and on time. This is because an untrapped orbit spans a $2\pi$-interval in $\theta$, while $\theta$ has to vary by $4\pi$ to complete a trapped orbit. Due to action conservation, after separatrix crossing the angle distribution remains nearly uniform over a $2\pi$-interval, and is nearly zero over the complementary $2\pi$-interval. Namely, the distribution function for trapped electrons reads,
\begin{equation}
\label{tilde}
\mathcal{F}_\gamma=\Theta[\theta-\theta_\alpha(I,t)]F_\alpha+\Theta[\theta-\theta_\beta(I,t)]F_\beta,
\end{equation}
where we have denoted $F_\alpha=f_\alpha(\mI_+)d_I\mI_+$ and $F_\beta=f_\beta(\mI_-)d_I\mI_-$, and where $\Theta(x)\approx 1$ whenever $0< x[4\pi]<2\pi$ and $\Theta(x)\approx 0$ whenever  $2\pi<x[4\pi]<4\pi$. As for $\theta_{\alpha,\beta}$, they are such that $d_t{\theta}_{\alpha,\beta}=d_IH$. From Eq.~(\ref{tilde}), it is clear that the time average of $\mathcal{F}_\gamma$ is $f_\gamma/2$, where $f_\gamma$ is given by Eq.~(\ref{df}) (see Ref.~\cite{DNC} for more details). Similarly, the mean value of $\mathcal{F}_\gamma$ over a large enough action interval, $\Delta I$, is independent of $\theta$, and is half of the averaged value of $f_\gamma$ over $\Delta I$. Now, the adiabatic limit corresponds to $\Delta I\rightarrow 0$. This means that, for trapped electrons,  when the initial action distribution function is smooth enough [i.e., when Eq.~(\ref{adia}) is fulfilled], 
\begin{equation}
\label{piege}
f(\theta,I,t) \approx f_\gamma(I)/2.
\end{equation}
Then, using $d\varphi dV=d\theta dI$, we find that the distribution function in variables $(\varphi,V)$, normalized to unity, is,
\begin{equation}
\label{piege2}
\tilde{f}(\varphi,V,t)\approx f_\gamma\left[I(\varphi,V,t)\right]/2.
\end{equation}

\subsubsection{Nonlinear frequency shift}
\label{IIC2}
As shown in Appendix~\ref{A}, for an EPW which is not externally driven, Poisson equation yields,
\begin{equation}
\label{disp}
-2\langle\cos(j\varphi)\rangle/j^2=\Phi_j,
\end{equation}
where $\langle . \rangle$ denotes statistical averaging (see Appendix~\ref{A}) and where,
\begin{equation}
\label{Phi}
\Phi\equiv\sum_{j\geq1}\Phi_j \cos(j\varphi)=k^2\phi/\omega_{pe}^2,
\end{equation}
$\omega_{pe}=\sqrt{n_ee^2/\varepsilon_0m}$ being the plasma frequency. Eq.~(\ref{disp}) is solved for given values of $k$ and $\Phi_1$. Hence, when $j=1$, this equation provides $V_\phi$ while, when $j\geq2$, it provides the amplitude of the harmonic, $\Phi_j$. More precisely, $\lcos$ as derived in the Appendix~\ref{B} is,
\begin{eqnarray}
\nonumber
\lcos&=&\int_{\mA_s}^{+\infty} \frac{\int_0^{\pi} \frac{\cos(j\varphi)d\varphi}{\sqrt{H(I)+\phi(\varphi)}}}  {\int_0^{\pi} \frac{d\varphi}{\sqrt{H(I)+\phi(\varphi)}}}   f_u(I)d(kI)   \\
\nonumber
&&+\int_{0}^{\mA_s}\frac{\int_0^{\varphi_{\max}} \frac{\cos(j\varphi)d\varphi}{\sqrt{H(I)+\phi(\varphi)}}}  {\int_0^{\varphi_{\max}} \frac{d\varphi}{\sqrt{H(I)+\phi(\varphi)}}}   f_\gamma(I)d(kI),\\
\label{cos}
\end{eqnarray}
where we have denoted $f_u(I)\equiv f_\alpha(I+V_\phi/k)+f_\beta(I-V_\phi/k)$. As for $H(I)$, it is derived from the definitions~Eqs.~(\ref{iua})-(\ref{it}) for the action. From Eq.~(\ref{cos}), it is clear that $\langle\cos(j\varphi)\rangle$ depends on $V_\phi$ and on the harmonic content of the potential. Consequently, one needs to solve Eq.~(\ref{disp}) self-consistently for $V_\phi$ and for the $\Phi_j$'s. In practice, we first calculate the $\Phi_j$'s, using a first guess for $V_\phi$, then calculate a new value for $V_\phi$ by solving Eq.~(\ref{disp}) when $j=1$, and iterate until convergence. 

Note that, from Eq.~(\ref{vphi}), $V_\phi$ is not the phase velocity, $\omega/k$. To derive $\omega$, following Ref.~\cite{lindberg}, we use the fact that the total electron momentum has to be conserved, since the wave has no magnetic field. Then,  $\langle dx/dt\rangle=0$, which yields,
\begin{equation}
\label{32}
\omega=-\left\langle\frac{d\varphi}{dt}\right\rangle.
\end{equation}
The derivation of $\langle d\varphi/dt\rangle$ is reported in the Appendix~\ref{B} where it is shown that,
\begin{equation}
\label{dphidt}
\left\langle \frac{d\varphi}{dt}\right\rangle=\int_{\mA_s}^{+\infty}\frac{\left[f_\alpha(I+V_\phi/k)-f_\beta(I-V_\phi/k)Ê\right]}{\pi^{-1}\int_{0}^{\pi} \frac{d\varphi}{\sqrt{2[H(I)+\phi(\varphi)]}}} d(kI).
\end{equation}
This concludes the derivation of the nonlinear wave frequency and harmonic content of the potential.

\subsection{Generalization to non-uniform plasma and wave amplitude}
\label{IID}
In this Paragraph, we explain how the results derived in Section~\ref{IIC} generalize  when the plasma and wave amplitude are space-dependent.

First of all, as shown in Appendix~\ref{A}, Eq.~(\ref{disp}) still holds. However, the wavenumber no longer remains constant. Indeed, from the very definition of $\bm{k}$ and $\omega$, $\partial_t\bm{k}=-\bm{\nabla}\omega$. The latter equation is actually solved along rays. Let, $\omega(\bm{x},t)\equiv \Omega_R[\bm{k}(\bm{x},t),\bm{x},t]$, where $\Omega_R$ solves the nonlinear dispersion relation. Then, the ray equations are,
\begin{eqnarray}
d_t\bm{x}_R&=&\partial_{\bm{k}}\Omega_R\vert_{\bm{x}}, \\
d_t\bm{k}_R&=&-\bm{\nabla}\Omega_R\vert_{\bm{k}}, 
\end{eqnarray}
where $\bm{k}_R(t)\equiv \bm{k}[\bm{x}_R(t),t]$. Hence, in two or three dimensions, the wavenumber direction usually changes, an effect that we specifically address in the Appendix~\ref{C}.

The distribution function and the dispersion relation are also calculated along rays. The change in the distribution function entailed by separatrix crossing is calculated as in Section~\ref{IIC1} (or, more generally, as in Ref.~\cite{benisti17I}), provided that one uses the convention,
\begin{eqnarray}
\label{36}
\frac{dV_\phi^*}{d\mA_s^*}\equiv\frac{dV_\phi^*/dt}{d\mA_s^*/dt},
\end{eqnarray}
where the derivatives are calculated along rays. Namely, in Eq.~(\ref{36}), $V_\phi^*\equiv V_\phi^*[\bm{x}_R(t),t]$ and $\mA_s^*\equiv \mA_s^*[\bm{x}_R(t),t]$. Note that, when the plasma is inhomogeneous, $V_\phi$ may change even when $\mA_s$ remains constant. 

Moreover, although Eqs.~(\ref{df2}) and (\ref{df3}) are still valid to account for the change in action due to separatrix crossing (when  $\vert d_tV_\phi/d_t\mA_s\vert<1$), one also needs to account for the fact that, due to inhomogeneity, $f_\alpha$ and $f_\beta$ vary slowly with time~\cite{benisti16},
\begin{equation}
\partial_tf_{\alpha,\beta}+k^{-1}\partial_IH_0\partial_xf_{\alpha,\beta}-\eta k^{-1}\partial_xH_0\partial_If_{\alpha,\beta}=0,
\end{equation}
where $H_0(x,I)$ is obtained by inverting Eq.~(\ref{igeneral}), and $\eta=1$ for $f_\alpha$ and $\eta=-1$ for $f_\beta$.

As regards the trapped orbits, and as discussed in Refs.~\cite{benisti16,dodin3}, there actually is one $f_\gamma(I)$ for each O-point, labeled by the O-point coordinate $\varphi_\text{O}$. Then, the change in $f_\gamma(I,\varphi_\text{O})$ is  only due to separatrix crossing, as given by Eq.~(\ref{df}) when $\vert d_tV_\phi/d_t\mA_s\vert<1$ or, more generally, as given in Ref.~\cite{benisti17I}.

Moreover, when the electric field is not exactly periodic, deriving the vector potential from Eq.~(\ref{32}) would let $A_0$ depend on space. Then, in a 3-D geometry, $\nabla\times A_0$ is not necessarily zero, so that the total electron momentum is not necessarily conserved, and Eq.~(\ref{32}) may actually not be exact. In order to derive $A_0$, we make use the following expansion, $A_0=A_0^{(0)}+\varepsilon^2A_0^{(2)}+\dots$, where $\varepsilon$ is of the order of the slow space or time variation of $A_0$. At zeroth order, $A_0^{(0)}$ clearly solves~Eq.~(\ref{32}). Hence, the from Amp\`ere-Maxwell law, it is such that $\langle\partial_{t^2}A_0^{(0)}\rangle=-(e/mk)\langle\partial_\varphi\phi\rangle$. Then, at second order, the Amp\`ere-Maxwell law yields,
\begin{equation}
\label{courant}
\nabla\times\left(\nabla\times A_0^{(0)}\right)=\mu_0j^{(2)},
\end{equation}
where the additional term $-c^{-2}\partial_{t^2}A^{(2)}$ has been dropped because it is of the order of $\varepsilon^4$, and where we have indentified $A_0^{(0)}$ and $j^{(2)}$ with their $\varphi$-averages. Hence, as the EPW grows, a magnetic field does build up and, at lowest order its expression is, $B_0=\nabla\times A_0$, where $A_0$ solves Eq.~(\ref{32}). Therefore, the EPW cannot be considered as purely electrostatic. Moreover, there is a mean current, given by Eq.~(\ref{courant}). These mean magnetic field and current have been observed numerically in Ref.~\cite{masson}, before the development of a Weibel-like instability. For times larger than the cyclotron period associated with $B_0$, the electron motion may be quite complicated (see Refs.~\cite{mit1,mit2,mit3,mit4} and references therein), but will not be discussed here. This would be way beyond the scope of this paper.  

\section{Nonlinear frequency shift for a wave growing in a uniform Maxwellian plasma}
\label{III}
In this Section, we explicitly calculate the values of the nonlinear frequency shift when the wave keeps growing in a uniform plasma, while its amplitude is also assumed to be uniform. Indeed, only in this case may one unambiguously derive $\delta \omega$ as a function of the amplitude of the first harmonic of the potential, $\Phi_1$.  Now, as shown in Ref.~\cite{benisti08}, such a calculation usually yields very accurate values for $\delta \omega$ even when the wave amplitude is not uniform, e.g., when the EPW is SRS-driven. 

Moreover, we assume here that the EPW grows in an  initially Maxwellian plasma, which is only possible if the wave is externally driven. Hence, the dispersion relation derived in Section~\ref{IIC2} may only be used here if the effect of the drive is negligible, which we henceforth assume. As discussed in Ref.~\cite{benisti08}, this assumption is valid for an SRS-driven plasma wave, and for conditions relevant to inertial confinement fusion, whenever $k\lambda_D\alt 0.35$. 

\subsection{Derivation of the nonlinear frequency shift}
\label{IVA}
\begin{figure}[!h]
\centerline{\includegraphics[width=8.6cm]{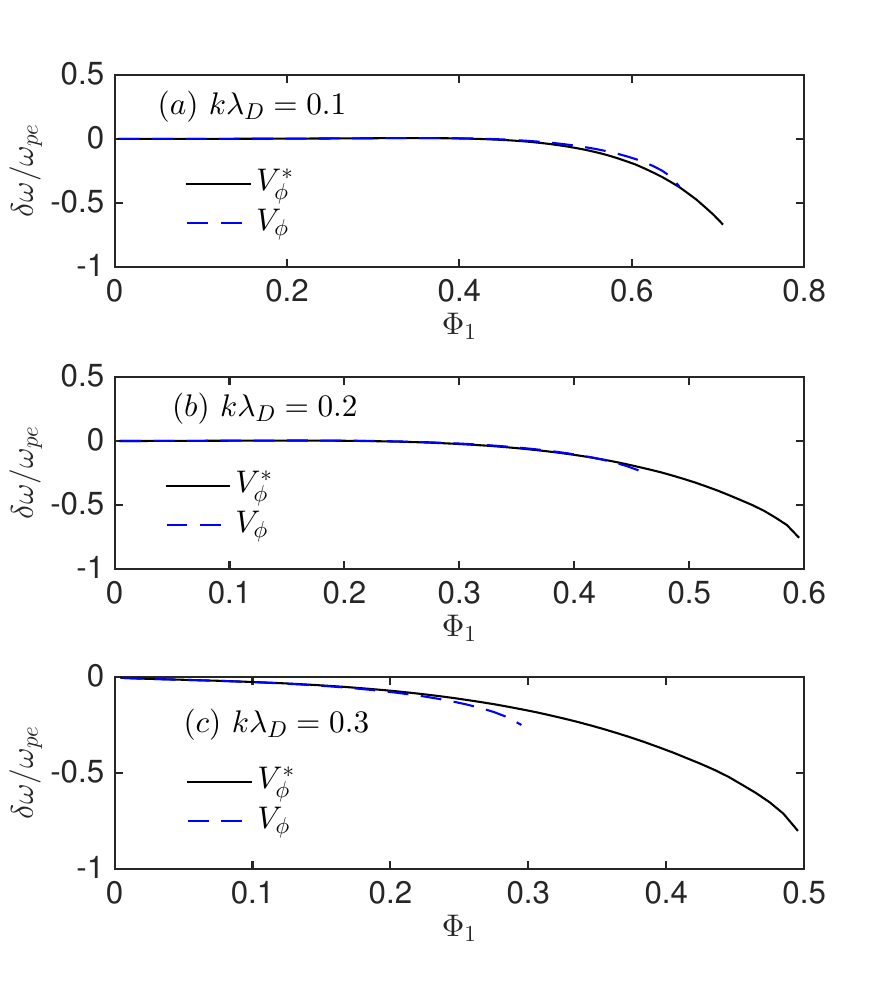}}
\caption{\label{f2} (Color online) Values of $\delta \omega$ as a function of the amplitude of the first harmonic of the potential when, panel (a): $k\lambda_D=0.1$ ; panel (b): $k\lambda_D=0.2$ ; panel (c): $k\lambda_D=0.3$.  The black solid line is for the exact adiabatic theory (with 3 harmonics), while the blue dashed line is obtained by replacing $V_\phi^*$ by the instantaneous and $I$-independent value, $V_\phi$, in Eq.~(\ref{df}).}
\end{figure}

Fig.~\ref{f2} shows the values derived for $\delta \omega$ as a function of $\Phi_1$, when $k\lambda_D=0.1$, $k\lambda_D=0.2$, and $k\lambda_D=0.3$.   For most values of $\Phi_1$, $\vert dV_\phi/d\mA_s\vert<1$ so that, $f_\alpha(I)=f_0(kI)=f_\beta(I)$ [because, for a Maxwellian plasma, $f_0(I)=f_0(-I)$]. As for $f_\gamma(I)$, it is derived from Eq.~(\ref{df}). 

Now, for the largest values of $\Phi_1$ in Fig.~\ref{f2}, $dV_\phi/d\mA_s<-1$. Then, an orbit leaving region $(\beta)$ may either go to region $(\gamma)$ or to region $(\alpha)$. For the corresponding values of $I$, the equality $f_\beta(I)=f_0(k_0I)$ still holds, but, as shown in Ref.~\cite{benisti17II},
\begin{eqnarray}
\label{38}
f_\alpha(I)&=&f_0[k_0I-2V_\phi^*(I)],\\
\label{39}
f_\gamma(I)&=&2f_0[k_0I-V_\phi^*(I)].
\end{eqnarray}

\begin{figure}[!h]
\centerline{\includegraphics[width=8.6cm]{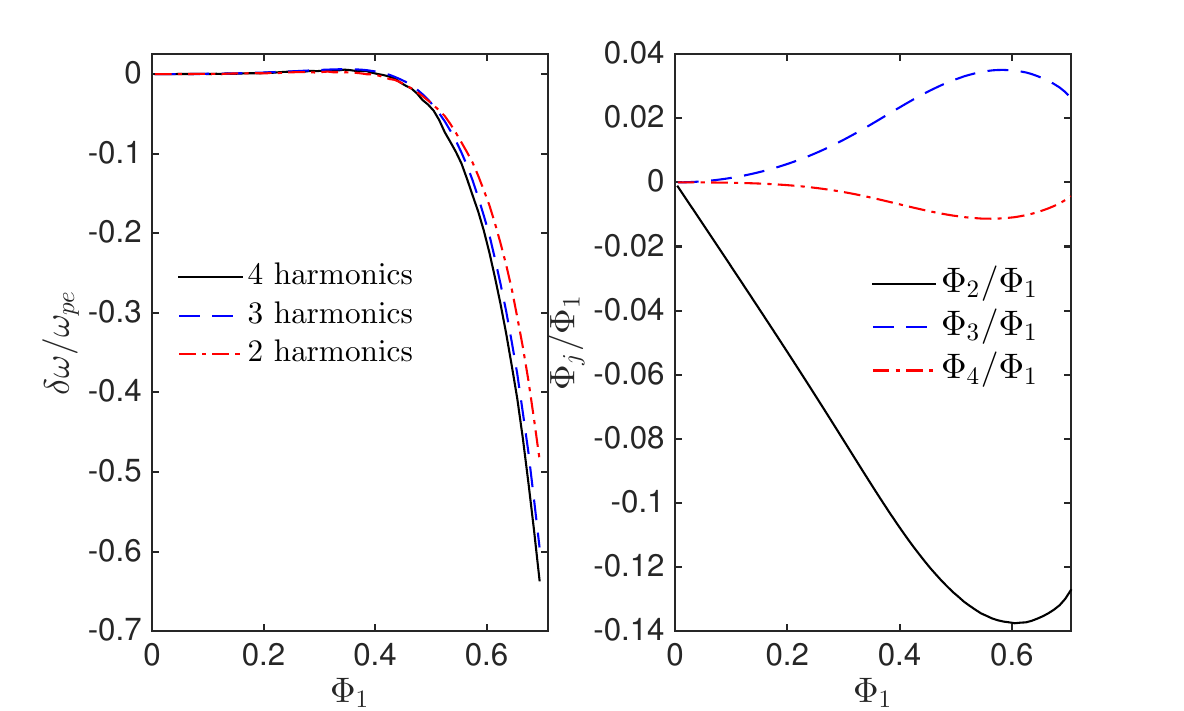}}
\caption{\label{f3} (Color online) Panel (a): values of $\delta \omega$ derived when $k\lambda_D=0.1$ and by accounting for 4 harmonics (black solid line), 3 harmonics (blue dashed line) and 2 harmonics (red dashed-dotted line) in the potential. Panel (b): relative magnitude of the harmonics amplitudes: $\Phi_2/\Phi_1$ (black solid line), $\Phi_3/\Phi_1$ (blue dashed line), $\Phi_4/\Phi_1$ (red dashed-dotted line)}
\end{figure}
Moreover, the derivation of $\delta \omega$ has always been performed by accounting for 3 harmonics in the potential $\Phi$, which is enough whenever $k\lambda_D\agt 0.1$. This is shown in Fig.~\ref{f3} for $k\lambda_D=0.1$, which corresponds to the most nonlinear case we studied because it allows for the largest values of $\Phi_1$.  From Fig.~\ref{f3}~(a), it is clear that the derivation of $\delta \omega$ with 3 harmonics has indeed converged, while Fig.~\ref{f3}~(b) shows that the amplitudes of $\Phi_j$, $j\geq2$, are always much smaller than that of $\Phi_1$. 
\subsection{Numerical check of the theoretical results}
We use test particles simulations in order to check the accuracy of our derivation for $\delta \omega$. Numerically, we solve the equations of motion derived from the Hamiltonian Eq.~(\ref{H}) when $\phi_1(t)=\phi_1(0)e^{\gamma t}$,
with $\phi_1(0)=10^{-5}$ and $\gamma=10^{-4}$. As for $V_\phi$ and the amplitudes of the second and third harmonics, their values are derived as a function of $\phi_1$ from our adiabatic calculation. Numerically we check that, for all $t$,
\begin{equation}
\label{num}
\frac{-2\langle\cos(j\varphi)\rangle_{\rm{num}}(t)}{j^2\Phi_j(t)}=1,
\end{equation}
where 
\begin{equation}
\langle\cos(j\varphi)\rangle_{\rm{num}}(t)=\sum_{l=1}^N w_l\cos[j\varphi_l(t)],
\end{equation}
the sum being over the $N$ electrons in the simulation ($N=4.8\times10^{6}$ in the case of Fig.~\ref{f4}). As for the weights $w_l$, they are $w_l=e^{-v_l^2/2}/(\sqrt{2\pi}n_v)$, where $v_l$ is the initial electron velocity (normalized to the thermal one) and $n_v$ is the number of electrons having the same initial velocity. 

\begin{figure}[!h]
\centerline{\includegraphics[width=8.6cm]{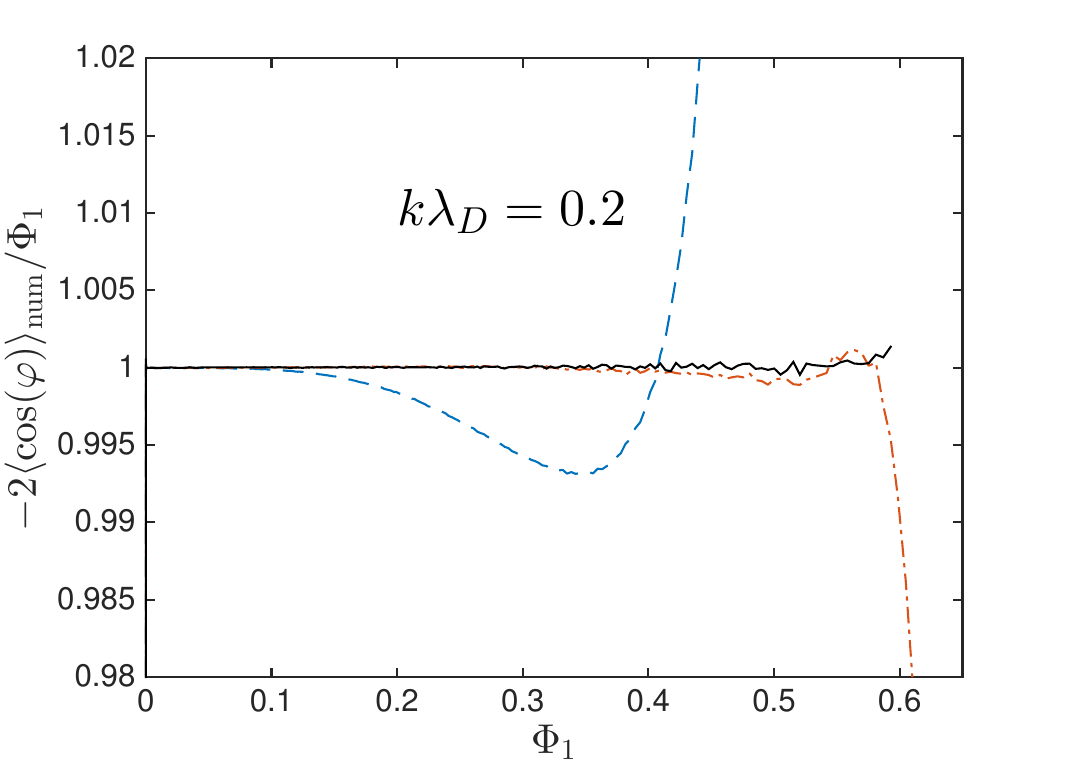}}
\caption{\label{f4} (Color online) Values of $-2\langle\cos(\varphi)\rangle_{\rm{num}}/\Phi_1$ as derived from test particles simulations. The black solid line is for  the exact adiabatic theory (with 3 harmonics), the blue dashed line is for the approximate theory when $V_\phi^*$ is replaced by $V_\phi$ in Eq.~(\ref{df}), while the red dashed-dotted line corresponds to the approximate theory assuming that separatrix crossing always leads to trapping.}
\end{figure}

Fig.~\ref{f4}, drawn for $k\lambda_D=0.2$, shows that Eq.~(\ref{num}) is very well fulfilled, thus evidencing the accuracy of the adiabatic predictions for $\omega$. Fig.~\ref{f4} also clearly shows deviations from Eq.~(\ref{num}) when $\delta \omega$ and the harmonics of the potential are derived by neglecting the $I$-dependence of $V_\phi^*$ in Eq.~(\ref{df}), or by assuming that separatrix crossing from regions~$(\alpha)$ or $(\beta)$ always entails trapping. 

\subsection{Simplifications to the general theory}
\label{IVB}
In this Paragraph, we mainly use the values derived for $\delta \omega$ in order to discuss when, and how, the general theory may be simplified.
\subsubsection{Importance of the harmonic content of the wave}
\label{IVB1}
\begin{figure}[!h]
\centerline{\includegraphics[width=8.6cm]{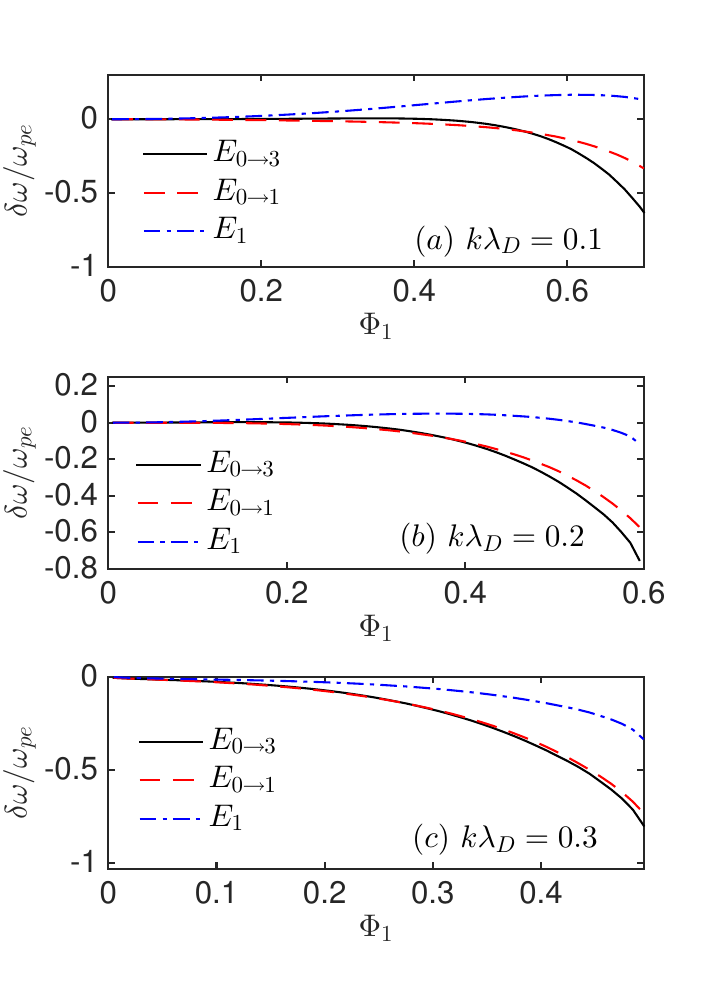}}
\caption{\label{f5} (Color online) Values of $\delta \omega$ as a function of the amplitude of the first harmonic of the potential when, panel (a): $k\lambda_D=0.1$ ; panel (b): $k\lambda_D=0.2$ ; panel (c): $k\lambda_D=0.3$.  The black solid line is obtained by accounting for harmonics 0 to 3 in the electric field, the red dashed results from a calculation including only the harmonics 0 and 1, while the blue dashed-dotted line is for a purely sinusoidal electric field.}
\end{figure}
Fig.~\ref{f5} compares the values found for $\delta \omega$ by accounting for harmonics 0 to 3 in the electric field with those obtained by accounting only for harmonics 0 and 1, or by assuming that the electric field is purely sinusoidal. This Figure clearly shows that the zeroth harmonic of the electric field (i.e., the vector potential), has to be accounted for in order to derive $\delta \omega$.  Neglecting $E_0$ may actually yield the wrong sign for $\delta \omega$, as may be seen in in Fig.~\ref{f5} when $k\lambda_D=0.1$ or $k\lambda_D=0.2$. Accounting only for $E_0$ and $E_1$ in the electric field yields quite accurate values for $\delta \omega$, except for the largest values of $\Phi_1$ when $k\lambda_D=0.1$. This is a very important result because, from Eqs.~(\ref{iua})-(\ref{it}) for the action, the derivation of the electron dynamics and distribution function is exactly the same when the electric field is purely sinusoidal as when it only includes the zeroth and first harmonics. Consequently, even for very large wave amplitudes, a sinusoidal derivation of $\langle\cos(\varphi)\rangle$ is still relevant. This is true within the adiabatic approximation, when one may use analytical results to derive $\langle\cos(\varphi)\rangle$ for a sinusoidal wave (see Ref.~\cite{benisti07}). As will be discussed in a forthcoming paper, this is also true when deriving the dispersion relation and distribution function without resorting to the adiabatic approximation. 

\subsubsection{Importance of using $V_\phi^*$}
\label{IVB2}
Fig.~\ref{f2} compares the values of $\delta \omega$ from the exact adiabatic formalism (with harmonics 0 to 3 in the potential), with the results obtained by replacing $V_\phi^*$ with the instantaneous and $I$-independent value, $V_\phi$, in Eq.~(\ref{df}). This approximation has been commonly used in the past, in particular in Refs.~\cite{dewar,krapchev,russell,lindberg}. When the plasma is homogeneous, it yields values for $\delta \omega$ which are quite accurate, as may seen in Fig.~\ref{f2}. However, the maximum value of $\Phi_1$ allowing a solution to the dispersion relation is smaller than for the exact theory. Actually it is all the smaller as $k\lambda_D$ increases. For $k\lambda_D=0.3$, $\max(\Phi_1)\approx 0.3$ for the approximate theory while $\max(\Phi_1)\approx 0.5$ for the exact one. Consequently, even when the plasma is homogeneous, one needs to use $V_\phi^*$, and not $V_\phi$, in the derivation of the distribution function which is, therefore, non-local in the wave phase velocity. 

Now, when the plasma is homogeneous and for the values of $k\lambda_D$ considered in Fig.~\ref{f2}, $V_\phi$ does not change much. This is the reason why, using either $V_\phi$ or $V_\phi^*$ in Eqs.~(\ref{df}) leads to nearly the same values for $\delta \omega$. However, if the plasma is inhomogeneous, $V_\phi$ may vary much more because, in addition to its nonlinear shift, it also changes with the plasma density or temperature. Then, replacing $V_\phi^*$ by $V_\phi$ in the derivation of the electron distribution function may lead to large errors, as discussed in Ref.~\cite{benisti17I}.

In conclusion, one needs to account for the non-locality in phase velocity of the electron distribution function, especially for large amplitude waves in a homogeneous plasma, or when the plasma is inhomogeneous. 

\subsubsection{Importance of the transition probabilities}
In most of the previous papers on the nonlinear frequency shift of an EPW, e.g. Refs.~\cite{dewar,krapchev,russell,lindberg}, it has been assumed that an increase in the wave amplitude systematically led to electron trapping. Consequently, $\delta \omega$ was derived by using Eq.~(\ref{df}) for $f_\gamma(I)$, together with $f_\alpha(I)=f_\beta(-I)=f_0(I)$. Fig.~\ref{f4} shows that these approximations are not too bad for a wave growing in a homogeneous plasma, because only for the largest amplitude is $\vert dV_\phi/d\mA_s\vert>1$. However, when the plasma is inhomogeneous, $V_\phi$ may vary due to a change in the plasma density or temperature, so that $\vert dV_\phi/d\mA_s\vert$ may be systematically larger than unity. This may actually entail interesting effects, such as a directed transport discussed in Ref.~\cite{leoncini}. Therefore, in general, one should use the correct transition probabilities, which leads to no additional technical complexity when deriving $\delta \omega$. 

\subsubsection{Importance of accounting for the change in $\bm{k}$-direction}
The importance of accounting for the change in the $\bm{k}$-direction in order to derive the electron distribution function is addressed in the Appendix~\ref{C}. In this Appendix, it is shown that a rotation in $\bm{k}$ enhances the fraction of trapped electrons, compared to the 1-D situation. However, this enhancement remains modest, and should be negligible, whenever $\bm{k}$ has rotated by less that about 30 degrees. For larger angles, no general conclusion may be drawn and any particular situation has to be analyzed along the same lines as in Appendix~\ref{C}. 
\section{Conclusion}
\label{IV}
In this paper, we addressed the nonlinear properties of adiabatic electron plasma waves, which play an important role in many areas of plasma physics, such as stimulated Raman scattering in fusion plasmas, or electron acceleration in the radiation belts~\cite{artemyev}.

We first specified which waves might be considered as adiabatic, and how to actually apply the adiabatic approximation. 

Then, we provided a self-consistent nonlinear theory for adiabatic EPW's, valid whatever their slow time and 3-D space variations. This theory led us to the following important results. 

We derived explicit values for the nonlinear frequency shift, $\delta \omega$, of a uniformly growing EPW. 

We showed that the nonlocality of the electron distribution function had to be accounted for in order to correctly estimate the limit, in amplitude, beyond which the dispersion relation could not be solved. This result is particularly important to derive the wave breaking limit, addressed in the companion paper~\cite{benisti20II}. 

We also showed that the vector potential of the EPW could not be neglected, except for small amplitudes. Moreover, we explicitly derived the magnetic field induced by an electron plasma wave, which has often been considered as purely electrostatic.

As regards the scalar potential, we showed that accounting for a single harmonic was often enough to correctly calculate the electron distribution. Actually, this result does not restrict to adiabatic plasma waves, and will be used in a forthcoming paper to address nonlinear EPW's when the adiabatic approximation is not valid. 

Finally, we addressed the impact, on the distribution function, of the wavenumber rotation, and showed that it enhanced electron trapping. The $\bm{k}$-rotation may be induced by plasma inhomogeneity or by an inhomogeneous frequency shift. The latter effect has been evidenced numerically in Ref.~\cite{yin}, and experimentally in Ref.~\cite{rousseaux}. In Ref.~\cite{yin}, it has been suspected to entail the EPW self-focussing together with the saturation of SRS. Moreover, together with the unstable growth of secondary modes, it provides a large extent to the EPW transverse spectrum, although it is not quite clear which effect actually dominates~\cite{rousseaux}. In the companion paper~\cite{benisti20II}, we address this issue theoretically. More precisely, from  the values computed in the present article for the frequency shift, we derive in Ref.~\cite{benisti20II} the transverse wavenumbers which only result from the inhomogenity in $\delta \omega$. This lets us unambiguously interpret  the Fourier spectra of the EPW that may be derived from particle in cell simulations, as in Ref.~\cite{rousseaux}.

In conclusion, this article provides a complete self-consistent theory for nonlinear adiabatic plasma waves which, we believe, was not available in previous publications. How to apply such a theory to actually describe nonlinear electron plasma waves and predict their evolution is an issue, which we explicitly address in the companion paper~\cite{benisti20II}. 
\appendix
\section{Dispersion relation at zeroth order in the space dependence of the fields}
\label{A}
\setcounter{equation}{0}
\newcounter{app}

\setcounter{app}{1}
Let us start with the expression Eq.~(\ref{E}) for the wave electric field, $E=E_0+\sum_{j\geq1} E_j \sin(j\varphi+\delta\varphi_j)$. Then, the charge density necessarily reads, 
\begin{equation}
\rho=\rho_0+\sum_{j\geq1} \left[\rho^c_j\cos(j\varphi+\delta\varphi_j)+\rho^s_j\sin(j\varphi+\delta\varphi_j) \right],
\end{equation}
where, as shown in Refs.~\cite{benisti07,benisti16},
\begin{equation}
\rho^c_j(\varphi)+i\rho^s_j(\varphi)=\frac{-ekn_e}{\pi} \int_{-\infty}^{+\infty}\int_{\varphi-\pi}^{\varphi+\pi}\tilde {f}(\varphi',V)e^{ij\varphi'}d\varphi' dV,
\end{equation}
$n_e$ being the electron density and $\tilde{f}$ the electron distribution function in variables $\varphi$ and $V$, normalized to unity.
Now, since we only want to derive the wave dispersion relation at zeroth order in the derivatives of the fields amplitudes, one may use for the potential, $\phi$, in the expression Eq.~(\ref{H}) for the Hamiltonian,
\begin{equation}
\label{A1}
\phi=\sum_{j\geq1}\phi_j\cos(j\varphi+\delta\varphi_j),
\end{equation}
where $\phi_j$ is given by Eq.~(\ref{phij}). Then, since the same phase shifts, $\delta \varphi_j$, appear in the expression for the field and for the potential, one may choose $\delta \varphi_j=0$. 

Moreover, from Gauss law, 
\begin{eqnarray}
\label{d1}
jkE_j&=&\rho^c_j /\varepsilon_0,\\
\label{d2}
\partial_xE_j&=&\rho^s_j/\varepsilon_0.
\end{eqnarray}
At zeroth order in the space derivative of $E_j$, Eq.~(\ref{d2}) reads,
\begin{equation}
\label{A2}
\rho^s_j=0.
\end{equation}
As shown in Ref.~\cite{benisti07}, Eq.~(\ref{A2}) is consistent with the neglect of the explicit $\theta$-dependence of the electron distribution function. Hence, one may use the approximation Eq.~(\ref{3}) for $F$. 

From Eqs.~(\ref{d1}) and~(\ref{phij}), the nonlinear EPW dispersion relation reads,
\begin{equation}
\label{A3}
-2\langle\cos(j\varphi)\rangle/j^2=\Phi_j,
\end{equation}
 where $\Phi_j=k^2\phi_j/\omega_{pe}^2$,  and where,
 \begin{equation}
 \label{A8}
\langle\cos(j\varphi)\rangle\equiv \frac{k}{2\pi}  \int_{-\infty}^{+\infty} \int_{-\pi}^{\pi}\tilde {f}(\varphi,V,t)\cos(j\varphi)d\varphi dV.
\end{equation}

\section{Statistical averaged values}
\label{B}
\setcounter{equation}{0}
\setcounter{app}{2}
In this Appendix, we explicitly derive the expressions for $\lcos$ and for $\langle d\varphi/dt\rangle$. 

Let us start with $\lcos$ which we derive from Eq.~(\ref{A8}) by making use of the canonical change of variables $(\varphi,V) \rightarrow (\theta,I)$. When $\varphi$ varies from $-\pi$ to $\pi$, it is clear that $\theta$ varies from $-\pi$ to $\pi$ in region $(\alpha)$, and from $\pi$ to $-\pi$ is region $(\beta)$. Then, using the fact that the Jacobian of the transformation is unity, one finds that the contribution to $\lcos$ from the untrapped electrons is, 
\begin{eqnarray}
\nonumber
\langle\cos(j\varphi)\rangle_u&=&\frac{1}{2\pi}\int_{\mA_s+V_\phi}^{+\infty}\int_{-\pi}^{\pi} f_\alpha(I)\cos(j\varphi)d(kI)d\theta +\\
\nonumber
&&\frac{1}{2\pi}\int_{\mA_s-V_\phi}^{+\infty}\int_{\pi}^{-\pi} f_\beta(I)\cos(j\varphi)d(kI)d\theta.
\end{eqnarray}
As regards the trapped electrons, there are 2 different values of $\theta$ for each value of $\varphi$. This translates the fact that, to one given value of $\varphi$ corresponds 2 different locations on the trapped orbit in phase space, one with $kV>V_\phi$ and one with $kV<V_\phi$. Consequently, when $\varphi$ varies from $-\pi$ to $\pi$, $\theta$ may either vary from  $-\pi$ to $\pi$ (on the branch of the trapped orbit corresponding to $kV>V_\phi$) or from $3\pi$ to $\pi$ (on the branch corresponding to $kV<V_\phi$). Then, using the approximation~Eq.~(\ref{piege}) for $f(\theta,I)$, one finds that the contribution to $\lcos$ from the trapped electrons is,
\begin{eqnarray}
\nonumber
\langle\cos(j\varphi)\rangle_t&=&\frac{1}{4\pi}\int_{0}^{\mA_s}\int_{-\pi}^{\pi} f_\gamma(I)\cos(j\varphi)d(kI)d\theta +\\
\nonumber
&&\frac{1}{4\pi}\int_{0}^{\mA_s}\int_{3\pi}^{\pi} f_\gamma(I)\cos(j\varphi)d(kI)d\theta.
\end{eqnarray}
Now, it is easy to show that,
\begin{equation}
\label{0B2}
d\theta=\pm\frac{\pi}{\int_{0}^{\varphi_{\max}} \frac{d\varphi'}{\sqrt{H(I)+\phi(\varphi')}}}\frac{d\varphi}{\sqrt{H(I)+\phi(\varphi')}},
\end{equation}
where $\varphi_{\max}>0$ is such that $H+\phi(\varphi_{\max})=0$ for a trapped orbit,  while $\varphi_{\max}=\pi$ for an untrapped orbit. Clearly, the plus sign is to be used in region $(\alpha)$ and the minus sign in region ($\beta$). Then, the change of variables $\theta\rightarrow\varphi$ yields,
\begin{equation}
\label{0B3}
\lcos_u=\int_{\mA_s}^{+\infty} \frac{\int_0^{\pi} \frac{\cos(j\varphi)d\varphi}{\sqrt{H(I)+\phi(\varphi)}}}  {\int_0^{\pi} \frac{d\varphi}{\sqrt{H(I)+\phi(\varphi)}}}   f_u(I)d(kI),
\end{equation}
where we have denoted $f_u(I)\equiv f_\alpha(I+V_\phi)+f_\beta(I-V_\phi)$. 
For trapped orbits, the plus sign in Eq.~(\ref{0B2}) is to be used when $\theta$ varies from $-\pi$ to $\pi$, and the minus sign when $\theta$ varies from $3\pi$ to $\pi$. Then,
\begin{equation}
\label{0B4}
\lcos_t=\int_{0}^{\mA_s}\frac{\int_0^{\varphi_{\max}} \frac{\cos(j\varphi)d\varphi}{\sqrt{H(I)+\phi(\varphi)}}}  {\int_0^{\varphi_{\max}} \frac{d\varphi}{\sqrt{H(I)+\phi(\varphi)}}}   f_\gamma(I)d(kI).
\end{equation}

Let us now derive $\langle d\varphi/dt\rangle$. Because frozen trapped orbits are symmetric with respect to $V=V_\phi$, they bring no contribution to $\langle d\varphi/dt\rangle$. Then, using exactly the same kind of calculation as for $\lcos$, one straightforwardly finds,
\begin{equation}
\left\langle \frac{d\varphi}{dt}\right\rangle=\int_{\mA_s}^{+\infty}\frac{\left[f_\alpha(I+V_\phi)-f_\beta(I-V_\phi)Ê\right]}{\pi^{-1}\int_{0}^{\pi} \frac{d\varphi}{\sqrt{2[H(I)+\phi(\varphi)]}}} d(kI).
\end{equation}

\section{Impact of the change in the wavenumber direction}
\label{C}
\setcounter{equation}{0}
\setcounter{app}{3}
In this Appendix, we address the 3-D situation when the direction of the wavenumber, $\bm{k}$, may slowly vary in space and time. We show how this impacts the electron distribution function, and discuss when the 1-D results remain essentially valid. 
 
 In 3-D, the action is still defined by~Eq.~(\ref{iu}) where, now, the change in $\varphi$ translates into a change in $x_r$, the local coordinate along the wavenumber. Hence, deriving the action requires writing the Hamiltonian for the wave-particle interaction in terms of $x_r$, and  in terms of the local coordinates across the wavenumber, which we denote by of $y_r$ and $z_r$. These coordinates read,
\begin{eqnarray}
\nonumber
x_r&=&x\cos\psi_r\cos\theta_r+y\cos\psi_r\sin\theta_r+z\sin\psi_r, \\
\nonumber
y_r&=&-x\sin\theta_r+y\cos\theta_r,\\
\nonumber
z_r&=&-x\sin\psi_r\cos\theta_r-y\sin\psi_r\sin\theta_r+z\cos\psi_r,
\end{eqnarray}
where $\theta_r$ varies between 0 and $2\pi$, and $\psi_r$ varies between $-\pi/2$ and $\pi/2$. Instead of considering $\psi_r$ and $\theta_r$ as functions of space and time, the analysis greatly simplifies if, for each electron, these angles are considered as given functions of time. Namely, $\theta_r(t)\equiv\theta_r[\bm{x}(t),t]$, $\psi_r(t)\equiv\psi_r[\bm{x}(t),t]$, where $\bm{x}(t)$ is the position of the considered electron, at time $t$. Then, introducing the generating function, $G=x_rv_{x_r}+y_rv_{y_r}+z_rv_{z_r}$, allows to derive the Hamiltonian for the conjugated variables $(x_r,y_r,z_r,v_{x_r},v_{y_r},v_{z_r})$,
\begin{equation}
\label{H3D}
H_{\text{3D}}=\frac{(\bm{v}+e\bm{A}_0/m)^2}{2}-\phi+\partial_tG.
\end{equation}
The extra term $\partial_tG$ couples the time variations of the various coordinates. This is essentially a geometrical effect entailed by the rotation of the wavenumber.  We proceed as in Section~\ref{IIB}, and perform the canonical change of variables, $(x_r,v_{x_r})\rightarrow(\varphi,V)$, that derives from the generating function $G'=\varphi V$. This yields the  3-D counterpart of Eq.~(\ref{H0}) for the Hamiltonian, 
\begin{eqnarray}
\nonumber
H_0&=&H-\frac{V_\phi^2}{2}+\frac{e^2A_0^2}{2m}+\frac{\left(\bm{v}_{\bot}+e\bm{A}_{\bot}/m\right)^2}{2}\\
\nonumber
&&-x_r(\dot{\theta}_rv_{y_r}\cos\psi_r+\dot{\psi}_rv_{z_r}) \\
\label{Hp0}
&&+\dot{\theta}_rz_rv_{y_r}\sin\psi_r-\dot{\theta}_ry_rv_{z_r}\sin\psi_r, 
\end{eqnarray}
where $\bm{v}_{\bot}$ and $\bm{A}_{\bot}$ are, respectively, the projections of the vector potential and electron velocity on the plane perpendicular to the wavenumber.  
In Eq.~(\ref{Hp0}), $H$ is still given by Eq.~(\ref{H}), while $V_\phi=\omega/k-eA_0/m-\dot{\theta}_r\cos\psi_ry_r-\dot{\psi}_rz_r$. Note that, along a ray where $y_r=z_r=0$,  $V_\phi$ is exactly the same as in 1-D. 

Let us now shift to action-angle variables. The action is still defined by Eq.~(\ref{iu}) where all the variables, except $\varphi$, are frozen. Hence, the action-angle variables are those of Hamiltonian $H$, which assumes the same expression as in 1-D. Consequently, in 3-D, we work with exactly the same action-angle variables as in 1-D.  When $\dot{\theta}_r$ and $\dot{\psi}_r$ are much less than $\omega$, the dynamics may be considered as slowly varying, so that $d_tI^{(\alpha,\beta)}\approx-\eta\left(\partial_{x_r}H_0\right)/k$, where $\eta=+1$ in region $(\alpha)$ of phase space and $\eta=-1$ in region $(\beta)$. Then, the time variations of the actions of untrapped orbits are coupled to those of the transverse velocities which, at first sight, leads to a situation much more complicated than in 1-D. In order to focus on the main effects induced by the change in the $\bf{k}$-direction, we restrict to a simplified two-dimensional ($\psi_r=0$) situation, when the $\varphi$-dependence of $\omega/k$ and of the amplitudes of the scalar and vector potentials are negligible. In this situation, $H_0$ may be reduced to,
\begin{equation}
\label{new1}
H_0=H-\frac{V_\phi^2}{2}-\dot{\theta}_rx_rv_{y_r}+\frac{v_{y_r}^2}{2},
\end{equation}
where $V_\phi=\omega/k-eA_0/m-\dot{\theta}_ry_r$, and where one may assume that $H$ only depends on $I$ and on $y_r$, through the $y_r$-dependence of $V_\phi$. For untrapped orbits, and from Eqs.~(\ref{iua})~and~(\ref{iub}), one finds $\partial H/\partial V_\phi=-(\eta\partial_IH)/k$. Then, Hamilton equations read,
\begin{eqnarray}
\nonumber
d_tI&\approx&-\eta\left(\partial_{x_r}H_0\right)/k\\
\label{B3}
&=&\eta \dot{\theta}_rv_{y_r}/k,Ê\\
\nonumber
d_tv_{y_r}&=&-\partial_{y_r}H_0\\
\nonumber
&=&-\partial_{V_\phi}H_0\partial_{y_r}V_\phi\\
\label{B4}
&=&-\dot{\theta}_r\left(\eta\partial_IH/k+V_\phi \right).
\end{eqnarray}
 Let us henceforth assume that all variables only depend on $\theta_r$ i.e., that $\theta_r$ may be used as a new time. This amounts to inverting the relation, $\theta_r=\theta_r(t)$ into, $t=t(\theta_r)$, which should always be possible, at least piecewise.  Moreover, let us now assume that the $\theta_r$-dependence of $k$ is negligible i.e., that the wavenumber mainly rotates. Then, introducing $J=\eta kI$, Eqs.~(\ref{B3})~and~(\ref{B4}) lead to,
\begin{eqnarray}
\label{Bn3}
d_{\theta_r}J&=&v_{y_r},\\
\label{Bn4}
d_{\theta_r}v_{y_r}&=&-(\partial_JH+V_\phi),
\end{eqnarray}
so that,
\begin{equation}
\label{new2}
d^2_{\theta_r}J=-(\partial_JH+V_\phi).
\end{equation}
Since the variations of $V_\phi$ are small, we henceforth neglect them, which lets us straightforwardly integrate Eq.~(\ref{new2}),
\begin{equation}
\label{B5}
\left(d_{\theta_r} J\right)^2=\left( J_0^2- J^2\right)\chi^2( J),
\end{equation}
where $ J_0$ is a constant (larger than $\vert J\vert$) and,
\begin{equation}
\chi( J)=\sqrt{1-[2H(J)-( J-V_\phi)^2]/( J_0^2- J^2)}.
\end{equation}
Eq.~(\ref{B5}) cannot be exactly solved without knowing the $\theta_r$-dependence of the wave amplitude, which depends on the particular physics problem which is addressed. However, one may note that  $\chi(J)$ is always less than unity. For a sinusoidal potential, $2H-( J-V_\phi)^2$ varies from 0 for a frozen orbit infinitely far from the separatrix to $(2-\pi^2/16)\phi_1$ when the frozen orbit is the separatrix. Then, $\theta'_r=\int_0^{\theta_r} \chi[J(u)]du< \theta_r$. Here, we consider the situation when, initially, the wave amplitude is infinitely small and the wavenumber is along the $x$-direction.  Hence, when $\theta_r\rightarrow0$, $J\rightarrow v_x(0)$, $d_{\theta_r} J\rightarrow v_{y}(0)$ and $\theta'_r/\theta_r\rightarrow 1$ (because, when the wave amplitude is infinitely small, all orbits may be considered infinitely far from the separatrix). Then, Eq.~(\ref{B5}) leads to, 
\begin{equation}
\label{B7}
J=v_{x}(0)\cos[\theta_r'(J)]+v_{y}(0)\sin[\theta_r'(J)].
\end{equation}
Now, for most untrapped orbits, $\theta'_r(J)\approx \theta_r$. Indeed, for a sinusoidal wave, $0.9<d_{\theta_r}\theta'_r<1$ whenever $\zeta\leq0.8$, where $\zeta=2\phi_1/(H+\phi_1)\leq1$.  Here, we are just pointing out the simple fact that, for most untrapped orbits, $J\approx v_{x_r}$, and Eq.~(\ref{B7}) with $\theta'_r$ replaced by $\theta_r$ just expresses the obvious result that $v_{x_r}=v_x\cos(\theta_r)+v_y\sin(\theta_r)$.

Let us now assume that the distribution functions of $v_x(0)$ and $v_y(0)$ are Maxwellians with, respectively, thermal velocities $v_{\rm{th}_x}$ and $v_{\rm{th}_y}$. From Eq.~(\ref{B7}), when $\cos(\theta_r)\neq 0$, the $J$-distribution function of untrapped electrons such that $\theta'_r\approx \theta_r$ is close to,
\begin{widetext}
\begin{equation}
\label{new3}
f_u(J)=\frac{1}{2\pi\vert\cos(\theta_r)\vert v_{\rm{th}_x}v_{th_y}}\int \exp\left(-\frac{v_y^2}{2v_{\rm{th}_y}^2} \right)\exp\left\{-\frac{[J-\sin(\theta_r)v_y]^2}{2v_{\rm{th}_x}^2\cos^2(\theta_r)} \right\}dv_y.
\end{equation}
\end{widetext}
When $\cos(\theta_r)=0$, $f_u(J)$ is obtained from Eq.~(\ref{new3}) by replacing $\cos(\theta_r)$ with unity, by inverting $v_{\rm{th}_x}$ and $v_{\rm{th}_y}$, and by replacing $v_x$ with $v_y$. Using the change of variables, $v=v_y/v_{\rm{th}_y}$ in Eq.~(\ref{new3}), one finds, after a few lines of calculation,
\begin{equation}
\label{B8}
f_u(J)=\frac{e^{-J^2/2v_{\rm{th}}^2}}{2\pi\vert\cos\theta_r\vert v_{\rm{th}_x}} \int e^{-\frac{v^2_{\rm{th}}}{2\cos^2\theta_rv^2_{\rm{th}_x}}\left[v-\frac{Jv_{\rm{th}_y}\sin\theta_r}{v^2_{\rm{th}}}Ê\right]^2}dv,
\end{equation}
where $v_{\rm{th}}=\sqrt{v_{\rm{th}_x}^2\cos^2\theta_r+v_{\rm{th}_y}^2\sin^2\theta_r}$. If the initial electron temperature is anisotropic, the number of untrapped electrons can change with $\theta_r$, only because of a change in the effective temperature. Moreover, while the EPW is growing with $\dot{\mA_s}>\vert\dot{V_\phi}\vert$ ($4\pi\mA_s$ being the area of the separatrix), electrons are being trapped when $\vert J-V_\phi\vert=\mA_s$, and they remain trapped unless $\mA_s$ decreases. Now, because $J$ depends on $\theta'_r$, as given by Eq.~(\ref{B7}), so does the condition on the initial electron velocity for trapping. Consequently, if $\bf{k}$ rotates, the EPW keeps on trapping new electrons, even when the increase in $\mA_s$ is insignificant. This would clearly not be the case if $\theta_r$ did not change. We conclude that the $\bf{k}$-rotation might lead to trapping enhancement, because it allows the EPW to keep on trapping new classes of electrons. More precisely, if the EPW keeps growing, with $\dot{\mA_s}>\vert\dot{V_\phi}\vert$,  an electron is untrapped at a given time $t^*$ if, whatever $t<t^*$,
\begin{equation}
\label{B0}
\left\vert \frac{ v_y(0)\sin[\theta'_r(t)-\theta'_r(t^*)]+J\cos\theta'_r(t)}{\cos\theta'_r(t^*)}  -V_\phi \right\vert>\mA_s[\theta'_r(t)],
\end{equation}
where, in Eq.~(\ref{B0}), $J\equiv J(t^*)$. For electrons such that $\theta'_r\approx \theta_r$, $f_u(J)$ is given by the left-hand side of Eq.~(\ref{B8}) where the integral is only over the values of $v=v_y(0)/v_{\rm{th}_y}$ fulfilling Eq.~(\ref{B0}). For the latter electrons,
\begin{equation}
\label{B9}
f_u(J)\leq\frac{e^{-J^2/2v_{\rm{th}}^2}}{\sqrt{2\pi}v_{\rm{th}}},
\end{equation}
the equality being reached when $\theta'_r$ remains contant i.e., when the $\bf{k}$-direction does not change. When the initial electron temperature is isotropic, Eq.~(\ref{B9}) clearly shows that the fraction of untrapped electrons is less than when the $\bf{k}$-direction remains constant. This evidences the trapping enhancement induced by the $\bf{k}$-rotation. If $\mA_s$ decreases in the time interval $[t_0,t^*]$, the electrons which are being detrapped are those such that Eq.~(\ref{B0}) is fulfilled at least once when $t$ varies from $t_0$ to $t^*$. Again, Eq.~(\ref{B9}) is recovered, the equality being reached if the  $\bf{k}$-direction remains constant or if $\mA_s[\theta'_r(t^*)]=0$.
\begin{figure}[!h]
\centerline{\includegraphics[width=8.6cm]{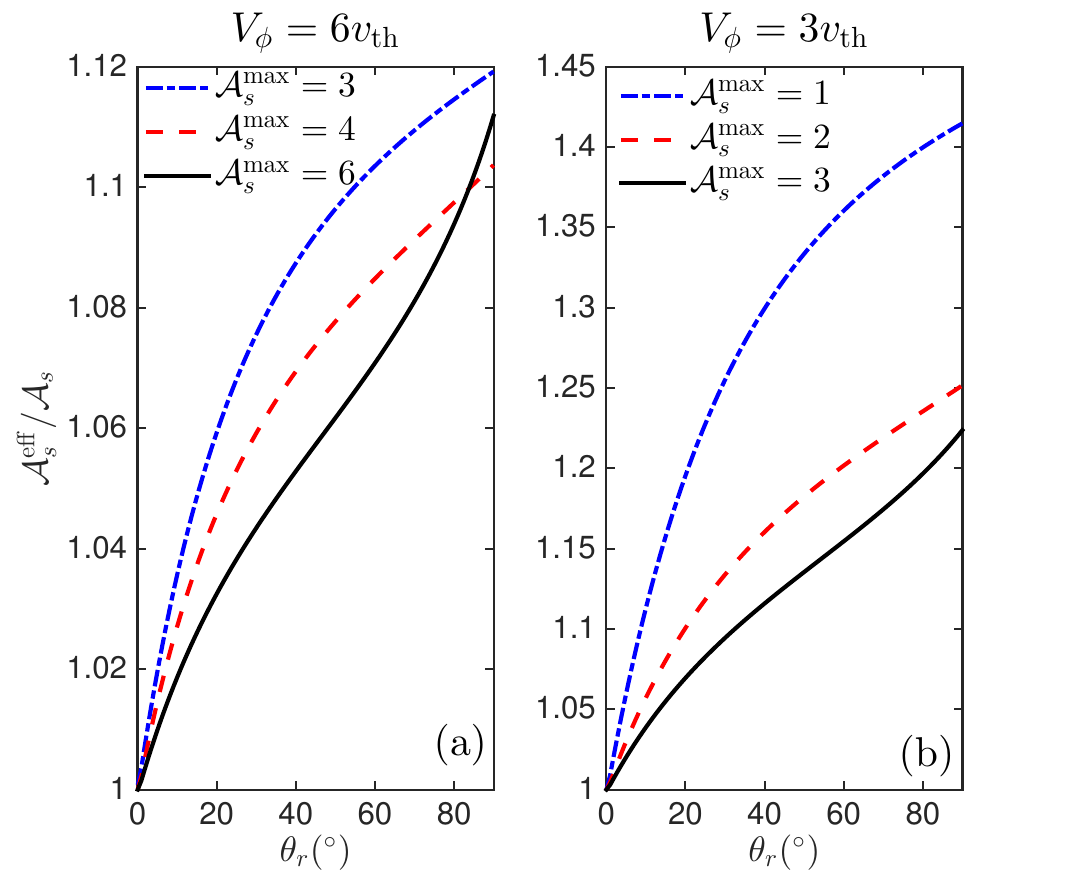}}
\caption{\label{fB1} (Color online) Values of $\mA_s^{\rm{eff}}/\mA_s$ when $V_\phi/v_{\rm{th}}=6$ [panel(a)], and $V_\phi/v_{\rm{th}}=3$ [panel (b)]. In panel (a), the blue dashed-dotted line is for $\mA_s^{\max}/v_{\rm{th}}=3$, the red dashed line is for $\mA_s^{\max}/v_{\rm{th}}=4$ and the black solid line is for $\mA_s^{\max}/v_{\rm{th}}=6$. In panel (b), the blue dashed-dotted line is for $\mA_s^{\max}/v_{\rm{th}}=1$, the red dashed line is for $\mA_s^{\max}/v_{\rm{th}}=2$ and the black solid line is for $\mA_s^{\max}/v_{\rm{th}}=3$. }
\end{figure}

How far from a Maxwellian $f_u(J)$  is, depends a lot on the physics situation which is addressed. In this Appendix, we mainly want to derive the conditions that make the 1-D approach of the main text valid, whatever the particular physics situation. To this end, we estimate the enhanced fraction of trapped electrons by making use of the approximation $\theta'_r=\theta_r$. Indeed, since $\theta'_r<\theta_r$, replacing $\theta'_r$ by $\theta_r$ in Eq.~(\ref{B7}) slightly overestimates the change in $J$ and, therefore, the change in the action distribution function. Moreover, for the sake of simplicity, we assume that the $\theta_r$-dependence of $\mA_s$ is the same for all electrons. When the $\bm{k}$-rotation is mainly due to the wavefront bowing entailed by the nonlinear frequency shift, $\mA_s$ first grows significantly before the direction of the wavenumber starts to change. In this Appendix, we consider $\theta_r$ variations of $\mA_s$ consistent with this particular physics situation, which we address in the companion paper~\cite{benisti20II}. More precisely, we assume that $\mA_s$ keeps growing with $\theta_r$ until it reaches its maximum value, $\mA_s^{\max}$. Moreover, we assume that $\mA_s$ first grows to $\mA_s^0$ ($0<\mA_s^0<\mA_s^{\max}$) when $\theta_r\approx 0$, and then smoothly varies with $\theta_r$. We checked that our results were essentially independent of $\mA_s^0$ and of the way $\mA_s$ smoothly varied with $\theta_r$. Fig.~\ref{fB1} corresponds to $\mA_s^0=\mA_s^{\max}/2$ and to a linear variation of $\mA_s$ with $\theta_r$ when $0\alt\theta_r\leq\max(\theta_r)$.

 In order to estimate the importance of trapping enhancement, we calculate the fraction of trapped electrons, $n_t=1-\int_{\vert J-V_\phi\vert\geq \mA_s}f_u(J)dJ$, as a function of $\theta_r$. To each value of $n_t$ we associate $\mA_s^{\rm{eff}}$ such that, 
\begin{equation}
n_t=\frac{1}{\sqrt{2\pi}v_{\rm{th}}}\int_{\vert J-V_\phi\vert\leq\mA_s^{\rm{eff}}} e^{-J^2/2v_{\rm{th}}^2}dJ.
\end{equation}
$\mA_s^{\rm{eff}}$ would be the value assumed by $\mA_s$ if the action distribution was a Maxwellian, i.e., if the wavenumber direction had remained constant. Because the $\bm{k}$-rotation enhances the electron trapping, $\mA_s^{\rm{eff}}\geq\mA_s$, as may be seen in Fig.~\ref{fB1}~(a) and (b) respectively when $V_\phi/v_{\rm{th}}=6$ and $V_\phi/v_{\rm{th}}=3$. These figures are plotted for three choices of $\mA_s^{\max}$, which varies from a value where nonlinear effects are small, to a value close to the wave breaking limit. As may be seen in Fig.~\ref{fB1}~(a), when $V_\phi/v_{\rm{th}}=6$,  $\mA_s^{\rm{eff}}$ does not exceed $\mA_s$ by more than about 10\%, even for angles as large as $90$\textdegree. We conclude that, when  $V_\phi/v_{\rm{th}}=6$, the $\bm{k}$-rotation does not significantly change the physics of the nonlinear wave-particle interaction, which is well captured by the 1-D approach of the main text. Fig.~\ref{fB1}~(b) shows that the same conclusion holds when $V_\phi/v_{\rm{th}}=3$, but only when $\theta_r\alt30$\textdegree. For larger angles, the need to account for the $\bm{k}$-rotation depends on the specific physics situation.

As regards the trapped electrons, their action is conserved and $H$ does not depend on $V_\phi$, so that the Hamilton equation for $v_{y_r}$ reads,
\begin{equation}
\label{B10}
d_tv_{y_r}=-\dot{\theta}_rV_\phi.
\end{equation}
For electrons close to the EPW ray, $V_\phi \approx \omega_{\rm{lin}}/k$ (since the nonlinear frequency shift always remain small compared to the linear wave frequency). Then, for those electrons, $\delta v_{y_r}Ê\approx -\theta_rV_\phi$. The wavenumber rotation entails a shift in the transverse velocity distribution and, consequently, the growth of a transverse component of the vector potential to mitigate it. Therefore, the change in the direction of the wavenumber leads to an additional growth for the magnetic field, compared to that already discussed in Section~\ref{IID}.


\begin{thebibliography}{10}
\bibitem{note}
Here, $T_B$ is for a sinusoidal wave because, when Eq.~(\ref{adia2}) is not fulfilled, the wave amplitude is so small that the wave is nearly sinusoidal.
\end{thebibliography}
\end{document}